\documentclass[preprint,aps,pre]{revtex4}
\usepackage{epsfig}
\usepackage{latexsym}
\begin{document}

\title{Breakdown of Universality in Directed Spiral Percolation}

\author{S. Sinha and S. B. Santra }

\affiliation{Department of Physics, Indian Institute of Technology
Guwahati,\\ Guwahati-781039, Assam, India.}

\date \today

\begin{abstract}
Directed spiral percolation (DSP), percolation under both directional
and rotational constraints, is studied on the triangular lattice in
two dimensions ($2D$). The results are compared with that of the $2D$
square lattice. Clusters generated in this model are generally
rarefied and have chiral dangling ends on both the square and
triangular lattices. It is found that the clusters are more compact
and less anisotropic on the triangular lattice than on the square
lattice. The elongation of the clusters is in a different direction
than the imposed directional constraint on both the lattices. The
values of some of the critical exponents and fractal dimension are
found considerably different on the two lattices. The DSP model then
exhibits a breakdown of universality in $2D$ between the square and
triangular lattices. The values of the critical exponents obtained for
the triangular lattice are not only different from that of the square
lattice but also different form other percolation models.
\end{abstract}
\maketitle

\section{Introduction}

A new site percolation model, directed spiral percolation (DSP), is
recently introduced by Santra\cite{dsp1,dsp2}. The DSP model is
constructed imposing both directional and rotational constraints on
the ordinary percolation (OP) model \cite{op}. The directional
constraint is in a fixed direction in space and the empty sites in
that direction are accessible to occupation. Due to the rotational
constraint the sites in the forward direction or in a rotational
direction, say clockwise, are accessible to occupation. The direction
of the rotational constraint is not fixed in space and it depends on
the direction from which the present site is occupied. Percolation
under only directional or only rotational constraints have been
studied independently and the corresponding models are known as
directed percolation (DP)\cite{dp} and spiral percolation
(SP)\cite{sp} respectively. It is already known that both DP and SP
models belong to different universality classes other than that of
OP. The DSP model is essentially a combination of DP and SP models and
it is constructed by imposing both the constraints simultaneously in
the same model. Recently, the DSP model has been studied on the square
lattice in $2$ dimensions ($2D$)\cite{dsp1,dsp2}. It has been found
that a new type of percolation cluster is generated in this
model. They are highly rarefied, anisotropic and chiral in nature. The
elongation of the clusters is in a different direction from the
imposed directional constraint. The values of the critical exponents
obtained are different from that of the OP, DP and SP. Consequently,
the DSP model belongs to a new universality class.

In this paper, the DSP model is studied on $2D$ triangular lattice and
the results are compared with that of the square lattice data. The
clusters are found more compact and less anisotropic on the triangular
lattice than on the square lattice. Most interestingly, it is found
that the values of the fractal dimension and some of the critical
exponents of the DSP model are considerably different on the square
and triangular lattices. Thus, there exists a breakdown of
universality in the DSP model when the results of the square and
triangular lattices are compared. In the following, the DSP model will
be described briefly. The results of the triangular lattice will be
then presented and compared with the square lattice data.

\section{The Model}
\begin{figure}
\bigskip
\centerline{\hfill \psfig{file=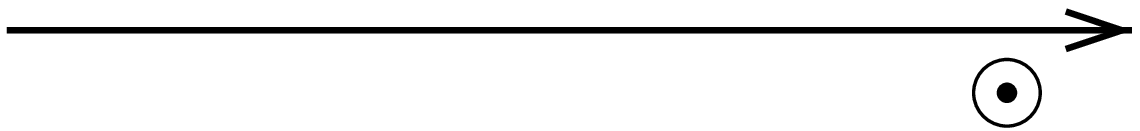,width=0.5\textwidth} \hfill}
\medskip
\centerline{\psfig{file=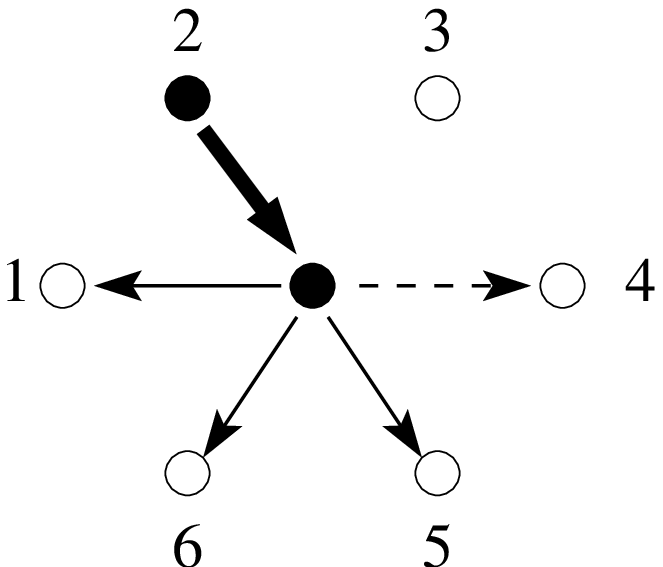,width=0.35\textwidth} }
\medskip
\centerline{\hfill \psfig{file=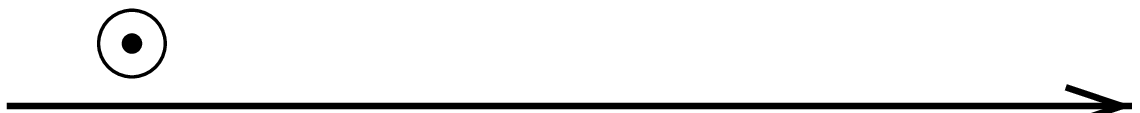,width=0.5\textwidth} \hfill}
\bigskip
\caption{\label{demo} Selection of eligible nearest neighbors for
 occupation of an already occupied site. Black circles are the
 occupied sites and open circles are the empty sites. Two thick long
 arrows from left to right represent the directional constraint. The
 presence of clockwise rotational constraint is shown by the encircled
 dots. The eligible nearest neighbors of the central occupied site
 will be selected here for occupation. Six nearest neighbours of the
 central site on the triangular lattice are marked as $1$ to $6$. The
 central site is occupied from the site $2$, marked by a thick
 arrow. Due to directional constraint, site $4$ on the right of the
 occupied site, is always eligible for occupation and it is indicated
 by a dotted arrow. Due to rotational constraint, sites $5$, $6$, and
 $1$ are eligible for occupation and they are indicated by thin
 arrows.}
\end{figure}

Detailed description of the model on the square lattice is given in
Ref.\cite{dsp1}.  A brief demonstration will be given here on the
triangular lattice. A left to right directional constraint and a
clockwise rotational constraint are imposed on the system defined on a
triangular lattice of size $L\times L$. Due to the directional
constraint an empty site on the right of an occupied site and due to
the rotational constraint the empty sites in the forward direction or
in the clockwise direction can be occupied. To generate clusters under
these two constraints a single cluster growth algorithm is developed
in Ref.\cite{dsp1} following the original algorithm of
Leath\cite{leath}. In this algorithm, the central site of the lattice
is occupied with unit probability. All six nearest neighbours of the
central site on the triangular lattice can be occupied with equal
probability $p$ in the first time step. As soon as a site is occupied,
the direction from which it is occupied is assigned to it. Selection
of empty nearest neighbour in the next MC time steps is illustrated in
Fig.\ref{demo}. Two long arrows from left to right in Fig.\ref{demo}
represent the directional constraint. The presence of the rotational
constraint is shown by the encircled dots. The black circles represent
the occupied sites and the open circles represent the empty sites. The
direction from which the central site is occupied is represented by a
short thick arrow. Now the nearest neighbours of the central occupied
site eligible for occupation will be identified. The dotted arrow
indicates the eligible empty site for occupation due to the
directional constraint and the thin arrows indicate the eligible empty
sites for occupation due to the rotational constraint. Since the
directional constraint is to the right, site $4$ is always eligible
for occupation. The rotational constraint acts in the forward or in
the clockwise direction with respect to the direction of approach to
the present occupied site. Since the central site is approached from
$2$, there are three sites, site $5$ in the forward direction and
sites $6$ and $1$ in the rotational direction, eligible for occupation
due to rotational constraint. Note that, on the square lattice only
two empty sites due to the rotational constraint were eligible for
occupation at any MC step. It is also important to note that, in this
model an occupied site can be reoccupied from a different direction
due to the rotational constraint\cite{dsp1}. A site is forbidden for
occupation from the same direction. On the triangular lattice, a site
then could be occupied at most $6$ times from $6$ different
directions. Due to the reoccupation of occupied sites, cluster
generation is time consuming in the rotationally constrained
models\cite{spst}. 

After selecting the eligible sites for occupation, they are occupied
with probability $p$. The coordinate of an occupied site in a cluster
is denoted by $(x$,$y)$. Periodic boundary conditions are applied in
both directions and the coordinates of the occupied sites are adjusted
accordingly whenever the boundary is crossed. At each time step the
span of the cluster in the $x$ and $y$ directions $L_x = x_{max} -
x_{min}$ and $L_y = y_{max} - y_{min}$ are determined. If $L_x$ or
$L_y\ge L$, the system size, then the cluster is considered to be a
spanning cluster. The critical percolation probability $p_c$ is
defined as below which there is no spanning cluster and at $p=p_c$ a
spanning cluster appears for the first time in the system.

\section{Results and Discussions} 
\begin{figure}
\bigskip
\centerline{\psfig{file=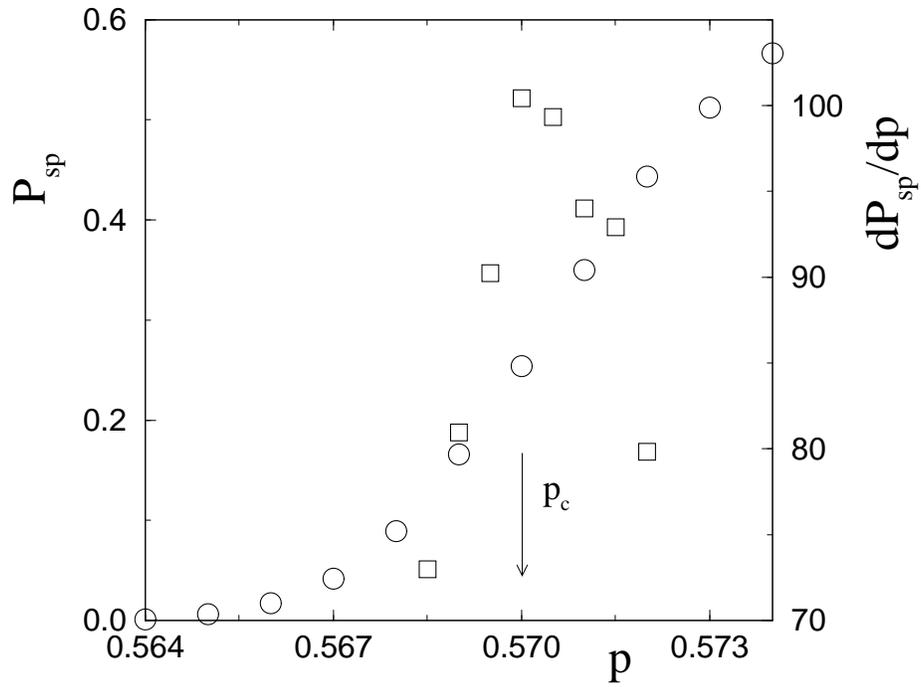,width=0.75\textwidth}}
\bigskip
\caption{\label{pcl} Plot of spanning probability $P_{sp}$ and the
slope $dP_{sp}/dp$ versus $p$. The circles represent $P_{sp}$ and the
squares represent the slope $dP_{sp}/dp$. The critical probability
$p_c$ is determined from the maximum slope. For the triangular
lattice, it is found that $p_c=0.5700\pm 0.0005$ as indicated by an
arrow.}
\end{figure}

Simulations are performed on the triangular lattice of several
different lattice sizes from $L=128$ to $L=2048$. The cluster size
distribution $P_s(p)$ is defined as $P_{s} = N_{s}/N_{tot}$ where
$N_s$ is the number of $s$-sited finite clusters in a total of
$N_{tot}$ clusters generated. The percolation threshold $p_c$ at which
a spanning cluster appears for the first time in the system is
determined by generating $N_{tot}=5\times 10^4$ clusters at different
site occupation probability $p$. The probability to have a spanning
cluster at a given site occupation probability $p$ is given by $P_{sp}
= n_{sp}/N_{tot} = 1 - \sum'_s{P}_s(p)$, where $n_{sp}$ is the number
of spanning clusters out of $N_{tot}$ clusters. The percolation
threshold $p_c$ is determined from the maximum slope
$(dP_{sp}/dp)_{max}$ of the curve $P_{sp}$ versus $p$. In
Fig.\ref{pcl}, $P_{sp}$ and $dP_{sp}/dp$ are plotted against $p$ for
$L=2048$. The percolation threshold $p_c$ could be identified as
$p_c=0.5700\pm 0.0005$ corresponding to the maximum slope. The
derivative is calculated using the central difference method for the
data points collected in an interval of $0.0005$.

\begin{figure}
\bigskip
\centerline{\psfig{file=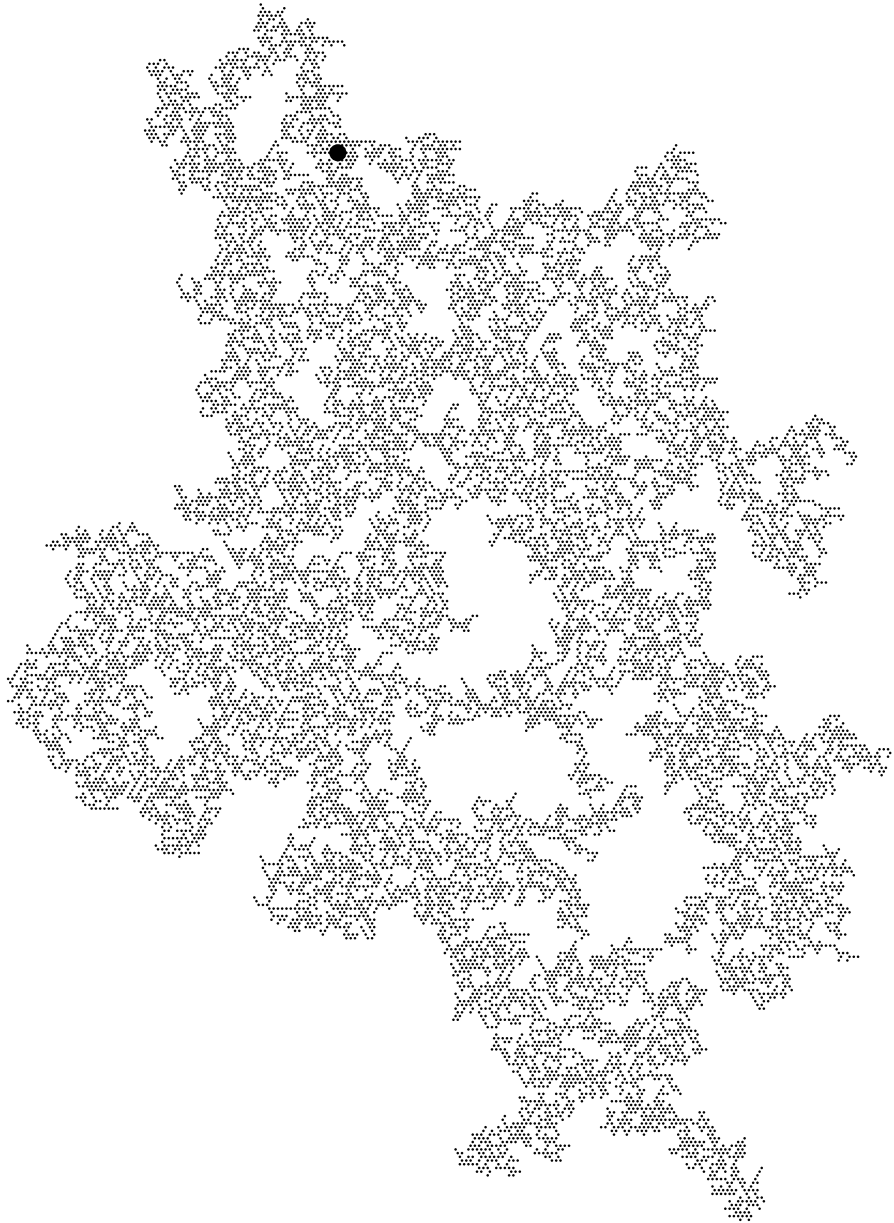,width=0.75\textwidth} }
\bigskip
\caption{\label{cluster} An infinite cluster on a $256\times 256$
triangular lattice at $p=0.5700$ is shown. The black dots are the
occupied sites. The solid black circle on the upper left corner is the
origin of the cluster. The cluster has holes of almost all possible
sizes. The elongation of the cluster is along the upper left to the
lower right diagonal and not along the imposed directional constraint
from left to right. The dangling ends are clockwisely rotated. }
\end{figure}

An infinite cluster generated on the triangular lattice of size
$L=256$ at $p_c=0.5700$ is shown in Fig.\ref{cluster}. The black dots
are the occupied sites and the solid black circle on the upper left
corner is the origin of the cluster. It could be seen that the
elongation of the spanning cluster is almost along the left upper to
the right lower diagonal of the lattice as it was seen on the square
lattice. In the case of charged particles, this is due to the
development of Hall voltage across the sample and perpendicular to the
applied in-plane electric field. As a result, an effective directional
field is developed along the left upper to the right lower diagonal of
the lattice. It has already been observed on the square lattice that
the DSP clusters are not merely the DP clusters in the presence of the
effective field. The DSP clusters contain features other than the DP
clusters. The clusters here contain holes of almost all possible sizes
and it has clockwisely rotated (chiral) dangling ends. It could also
be noticed that the infinite cluster on the triangular lattice is more
compact and less anisotropic in comparison to the infinite cluster on
the square lattice.

\begin{figure}
\bigskip
\centerline{\hfill \psfig{file=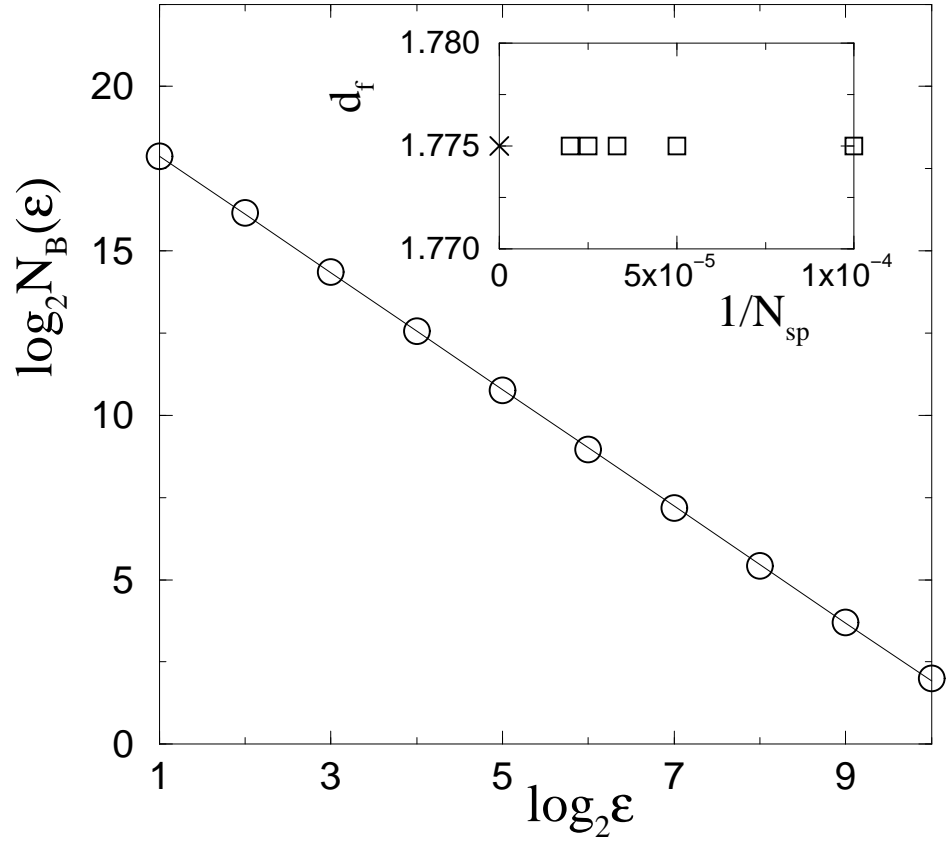,width=0.75\textwidth} \hfill}
\bigskip
\caption{\label{fracd} Number of boxes $N_B(\epsilon)$ is plotted
against the box size $\epsilon$. Data are averaged over $50,000$
spanning clusters generated on a triangular lattice of size
$L=2048$. The fractal dimension is found $d_f=1.775\pm 0.004$. In the
inset, $d_f$ is plotted against $1/N_{sp}$, the number of spanning
clusters. It could be seen that the value has converged with the
realizations.}
\end{figure}

The fractal dimension $d_f$ of the infinite clusters at $p_c=0.5700$
on the triangular lattice of size $L=2048$ is determined by the box
counting method. The number of boxes $N_B(\epsilon)$ is expected to
grow with the box size $\epsilon$ as $N_B(\epsilon) \sim
\epsilon^{d_f}$ where $d_f$ is the fractal dimension. In
Fig. \ref{fracd}, $N_B(\epsilon)$ is plotted against the box size
$\epsilon$. The data are averaged over $5\times 10^4$ spanning
clusters. A reasonably good straight line is obtained in the
$\log-\log$ scale. The fractal dimension is found $d_f= 1.775\pm
0.004$. The error is due to the least square fitting of the data
points taking into account the statistical error of each point. In
order to check the convergence of the value of the fractal dimension,
$d_f$ is plotted against $1/N_{sp}$ in the inset of
Fig. \ref{fracd}. It could be seen that the value of $d_f$ remains
unchanged over $10^4$ to $5\times 10^4$ spanning clusters. The value
of $d_f$ has also been estimated from finite size (FS) scaling
$S_\infty \sim L^{d_f}$, where $S_\infty$ is the size of the largest
cluster at $p=p_c$. The lattice size $L$ changes from $2^7$ to
$2^{11}$. It is found that $d_f(FS) = 1.80\pm 0.03$, which is within
the error bar of the other estimate. The fractal dimension $d_f$
obtained here is higher than that of $d_f\approx 1.733$\cite{dsp1}
($d_f(FS) \approx 1.72$ \cite{dsp2}) on the square lattice. Also
notice that the value of $d_f$ obtained here is smaller than the
fractal dimensions obtained in OP ($91/48$ \cite{nn}) and SP ($1.969$
\cite{hede}) and it is slightly higher than DP ($1.765$
\cite{spst}). Vacancies are generated into the cluster as it grows. At
the same time, due to the higher number of branching on the triangular
lattice the cluster penetrate into itself more and more than on the
square lattice. As a result, the infinite clusters are less rarefied
on the triangular lattice than on the square lattice.

\begin{figure}
\bigskip
\centerline{\hfill \psfig{file=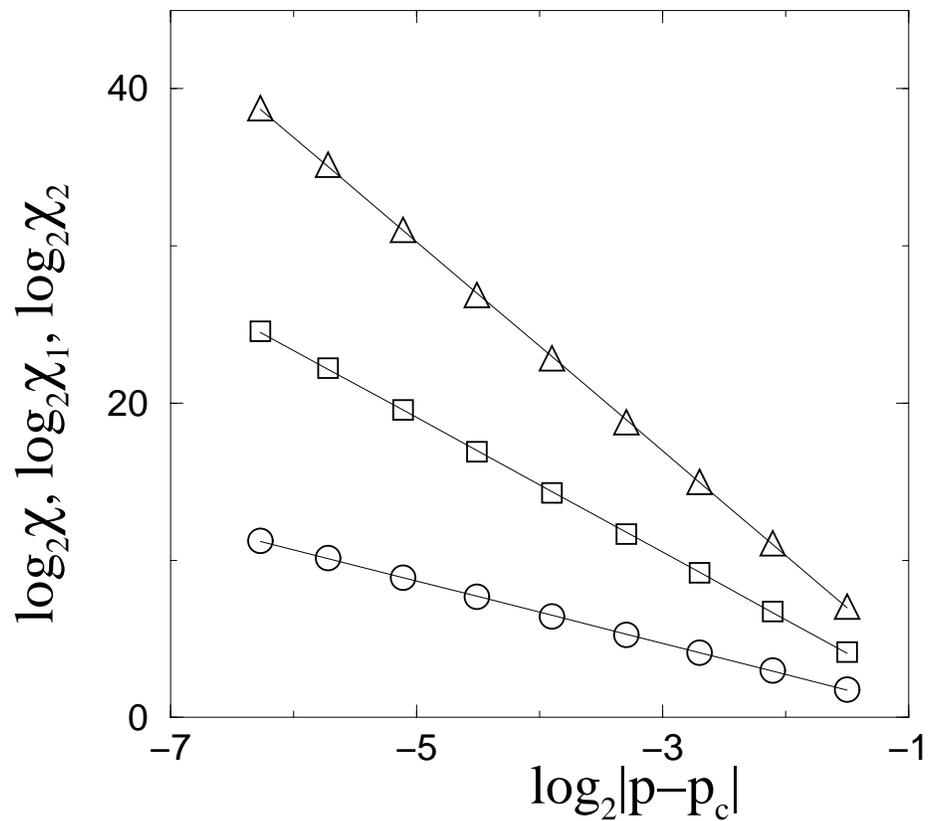,width=0.75\textwidth} \hfill}
\bigskip
\caption{\label{chi123} Plot of the first, second and third moments
$\chi$, $\chi_1$, and $\chi_2$ of the cluster size distribution versus
$|p-p_c|$ for a triangular lattice of size $L=2048$. Different symbols
are: circles for $\chi$, squares for $\chi_1$, and triangles for
$\chi_2$. The solid lines represent the best fitted straight lines
through the data points. The corresponding critical exponents are
found as $\gamma=1.98\pm 0.02$, $\delta= 4.30\pm 0.04$, and $\eta=
6.66\pm 0.08$. }
\end{figure}

\begin{figure}
\bigskip
\centerline{\hfill \psfig{file=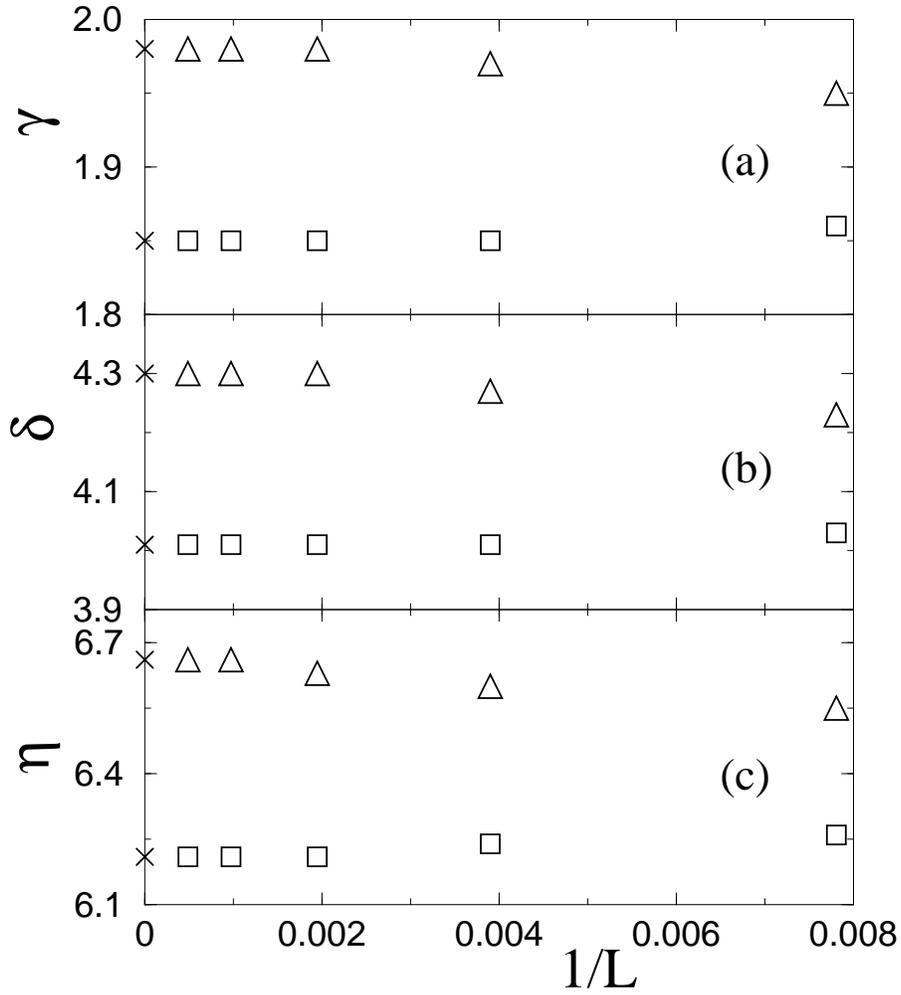,width=0.75\textwidth} \hfill}
\bigskip
\caption{\label{mexp} Plot of the exponents $\gamma$, $\delta$ and
$\eta$ against the inverse system size $1/L$. The system sizes
considered are: $L=128,256,512,1024$ and $2048$. The squares represent
the square lattice data and the triangles represent the triangular
lattice data. The exponents are extrapolated upto $L\rightarrow\infty$
and the extrapolated values are marked by crosses. It can be seen that
the exponents are significantly different on the two lattices. }
\end{figure}

Since the fractal dimension $d_f$ is different from that of the square
lattice value, it is then expected that the values of the other
critical exponents will also be different form that of the square
lattice in order to satisfy the scaling relations among the critical
exponents. The critical exponents related to the different moments of
the cluster size distribution $P_s(p)$ are now estimated. The scaling
function form of the cluster size distribution $P_s(p)$ for single
cluster growth technique, in which the central site is occupied with
unit probability, is assumed to be
\begin{equation}
\label{scalef}
P_s(p)=s^{-\tau+1}{\sf f}[s^\sigma(p-p_c)]
\end{equation}
where $\tau$ and $\sigma$ are two exponents. The assumed form of the
scaling function $P_s(p)$ is the same as that of the square
lattice. The first moment $\chi = \sum'_s sP_s(p)$ corresponds to the
average cluster size. Next two higher moments are defined as $\chi_1 =
\sum'_s s^2P_s(p)$ and $\chi_2 = \sum'_s s^3P_s(p)$. The primed sum
represents the sum over all the finite clusters.  As $p\rightarrow
p_c$, the moments $\chi$, $\chi_1$, and $\chi_2$ of $P_s(p)$ become
singular with their respective critical exponent $\gamma$, $\delta$,
and $\eta$. The critical exponents are defined as
\begin{equation}
\label{expns}
\hfill \chi \sim |p-p_c|^{-\gamma},\hspace{0.2cm} \chi_1 \sim
|p-p_c|^{-\delta}, \hspace{0.2cm} \& \hspace{0.2cm} \chi_2 \sim
|p-p_c|^{-\eta}. \hfill
\end{equation}
To estimate the values of $\gamma$, $\delta$ and $\eta$ on the
triangular lattice, the average cluster size $\chi$ and two other
higher moments $\chi_1$ and $\chi_2$ are measured generating $5\times
10^4$ finite clusters below $p_c$ for different $p$ values on several
lattice sizes. In Fig.\ref{chi123}, $\chi$, $\chi_1$ and $\chi_2$ are
plotted against $|p-p_c|$ for the system size $L=2048$. The circles
represent $\chi$, the squares represent $\chi_1$ and the triangles
represent $\chi_2$. The values of the exponents obtained are $\gamma =
1.98 \pm 0.01$, $\delta = 4.30 \pm 0.02$ and $\eta = 6.66 \pm 0.04$
for $L=2048$. The errors quoted here are the standard least square fit
error taking into account the statistical error of each single data
point. Because of the error bar $\Delta p_c=0.0005$ in the threshold,
all the exponents have also been estimated for two other critical
probabilities $p_c\pm\Delta p_c$. The values of the exponents obtained
for $p=0.5695$ are $\gamma \approx 1.96$, $\delta \approx 4.27$, and
$\eta \approx 6.60$ whereas for $p=0.5705$ they are $\gamma \approx
2.00$, $\delta \approx 4.33$ and $\eta \approx 6.71$. The values of
the critical exponents are then taken as: $\gamma = 1.98 \pm 0.02$,
$\delta = 4.30 \pm 0.04$ and $\eta = 6.66 \pm 0.08$. The values of the
critical exponents at the optional thresholds $p_c\pm\Delta p_c$ are
now within error bars. A comparison of the values of the critical
exponents obtained on the triangular and square lattices is made in
Fig.\ref{mexp} for several lattice sizes. In Fig.\ref{mexp}, the
values of $\gamma$, $\delta$ and $\eta$ are plotted against the
inverse system size $1/L$. The squares represent the square lattice
data and the triangles represent the triangular lattice data. The data
for the square lattice is taken from Ref.\cite{dsp1} except for
$L=2048$. In Ref.\cite{dsp1}, data were reported upto the maximum
lattice size $L=1024$. For the sake of comparison with $L=2048$
triangular lattice data, new estimates of the critical exponents have
also been made on $L=2048$ square lattice. The results of $L=2048$
square lattice are in good agreement with that of the smaller system
sizes. On the square lattice of size $L=2048$, the values of the
critical exponent obtained are: $\gamma = 1.85 \pm 0.01$, $\delta =
4.01 \pm 0.04$ and $\eta = 6.21 \pm 0.08$. The results on both the
lattices are then extrapolated upto $L\rightarrow \infty$, the
infinite system size. The extrapolated values of the exponents are
marked by crosses. The values of the critical exponents are found very
different (beyond the error bars) on the square and triangular
lattices. The triangular lattice values of the exponents are higher
than that of the square lattice. The clusters then grow much larger in
size on the triangular lattice than on the square lattice for a given
$p$. This might be due to higher number of branching possibilities on
the triangular lattice. Also notice that the value of $2\delta -
\gamma = 6.62$ is very close to the value of the exponent
$\eta=6.66$. The values of the exponents then satisfy the scaling
relation $\eta = 2\delta - \gamma$ \cite{dsp1} within error bars. The
exponents are not only different from the square lattice values but
also different from that of other percolation models, OP
\cite{nn,dsp1}, DP \cite{ebg,dsp1} and SP \cite{sp,spst,dsp1}.

\begin{figure}
\bigskip
\centerline{\hfill \psfig{file=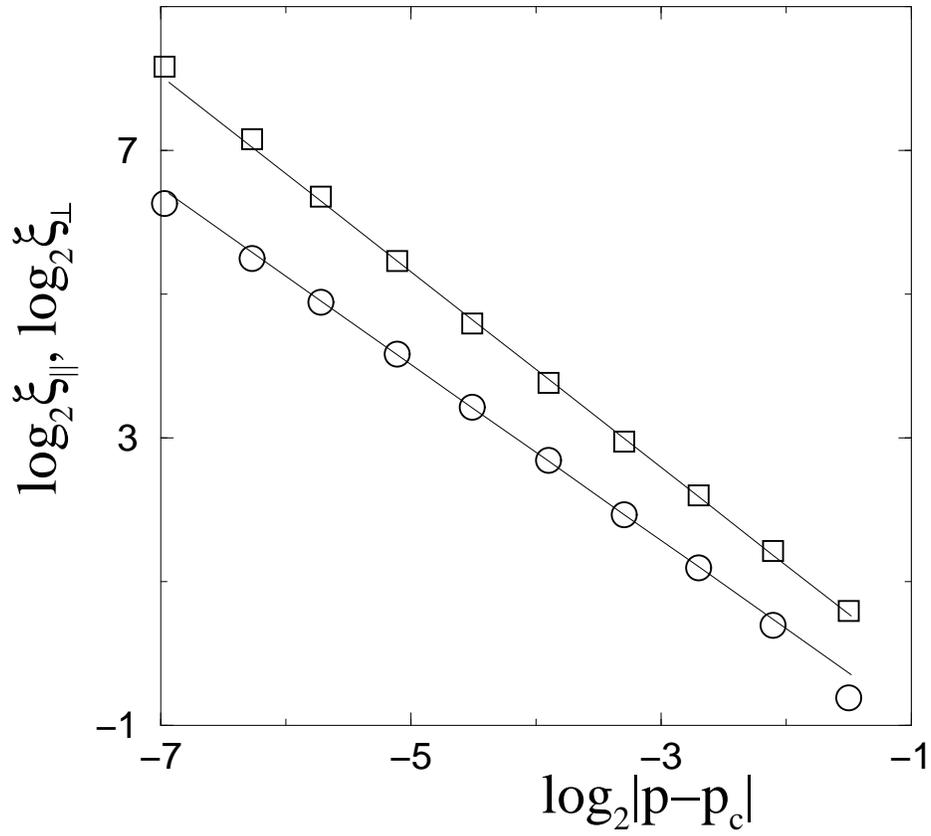,width=0.75\textwidth} \hfill}
\bigskip
\caption{\label{corrl} The connectivity lengths, $\xi_\parallel$ and
$\xi_\perp$, are plotted against $|p-p_c|$ for a triangular lattice of
size $L=2048$. The circles represent $\xi_\perp$ and the squares
represent $\xi_\parallel$. The solid lines represent the best fitted
lines through the data points. The critical exponents are found as
$\nu_\parallel= 1.36\pm 0.02$ and $\nu_\perp = 1.23 \pm 0.02$. }
\end{figure}

\begin{figure}
\bigskip
\centerline{\hfill \psfig{file=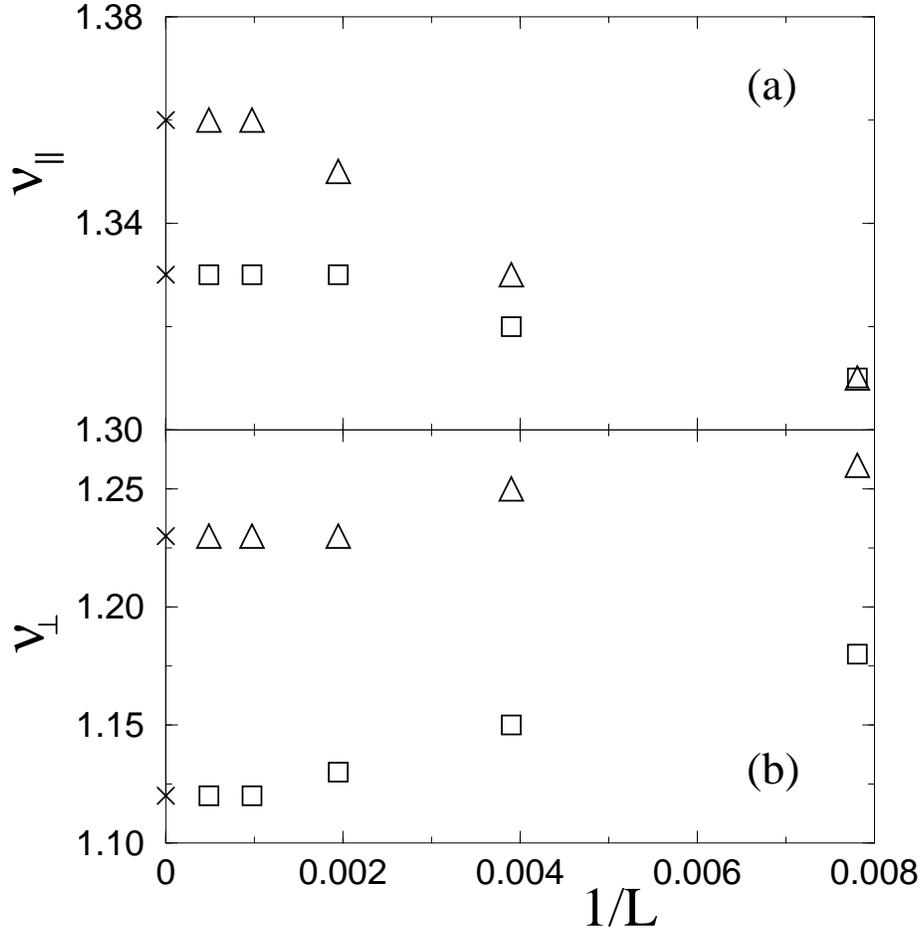,width=0.75\textwidth} \hfill}
\bigskip
\caption{\label{nuplpp} The connectivity exponents $\nu_\parallel$ and
$\nu_\perp$ are plotted against the inverse system size $1/L$ for the
square and triangular lattices. The system size changes form $L=128$
to $2048$ as in Fig.$6$. The squares represent the square lattice data
and the triangles represent the triangular lattice data. Extrapolated
values to the infinite system size ($1/L=0$) are marked by
crosses. The value of $\nu_\perp$ seems to be different on the two
lattices whereas $\nu_\parallel$ is close to the square lattice
value. }
\end{figure}

There are two connectivity lengths, $\xi_\parallel$ and $\xi_\perp$,
for the anisotropic clusters. Here, $\xi_\parallel$ is along the
elongation of the cluster and $\xi_\perp$ is along the perpendicular
direction to the elongation. The connectivity lengths are defined as
$\xi_\parallel^2=2\sum'_sR^2_\parallel sP_s(p)/\sum'_ssP_s(p)$
and $\xi_\perp^2=2\sum'_sR^2_\perp sP_s(p)/\sum'_ssP_s(p)$
where $R_\parallel$ and $R_\perp$ are the radii of gyration with
respect to two principal axes of the cluster. They are estimated from
the eigenvalues of the moment of inertia tensor, a $2\times 2$ matrix
here. $\xi_\parallel$ and $\xi_\perp$ diverge with two different
critical exponents $\nu_\parallel$ and $\nu_\perp$ as $p\rightarrow
p_c$. The critical exponents $\nu_\parallel$ and $\nu_\perp$ are
defined as
\begin{equation}
\label{expnc}
\xi_\parallel\sim |p-p_c|^{-\nu_\parallel} \hspace{0.5cm} \& \hspace
{0.5cm} \xi_\perp\sim |p-p_c|^{-\nu_\perp}.
\end{equation}
The connectivity lengths, $\xi_\parallel$ and $\xi_\perp$, for the
system size $L=2048$ are plotted against $|p-p_c|$ in
Fig.\ref{corrl}. Data are averaged over $5\times 10^4$ clusters. The
squares represent $\xi_\parallel$ and the circles represent
$\xi_\perp$. The corresponding exponents are $\nu_\parallel = 1.36\pm
0.02$ and $\nu_\perp = 1.23\pm 0.02$. The errors quoted here are the
least square fit errors. The values of the exponents are also estimated
at $p_c\pm\Delta p_c$. For $p=0.5695$, the values obtained are
$\nu_\parallel \approx 1.35$ and $\nu_\perp \approx 1.21$ and for
$p=0.5705$, the values obtained are $\nu_\parallel \approx 1.37$ and
$\nu_\perp \approx 1.24$. There is a little variation and the values
of the critical exponents are approximated as: $\nu_\parallel =
1.36\pm 0.02$ and $\nu_\perp = 1.23\pm 0.02$. The error bars now
include the values of $\nu_\parallel$ and $\nu_\perp$ at the optional
$p_c$s. To compare the square and triangular lattice data, simulations
have been performed on other smaller system sizes. In
Fig. \ref{nuplpp}, the exponents $\nu_\parallel$ and $\nu_\perp$ are
plotted against the inverse system sizes $1/L$ for both the square and
triangular lattices. The squares represent the square lattice data and
the triangles represent the triangular lattice data. Data of the
square lattice is taken from Ref.\cite{dsp1} except for $L=2048$. New
estimates of the exponents are also made on $L=2048$ square
lattice. The exponents are extrapolated upto $L\rightarrow \infty$ and
they are marked by crosses. Notice that, the exponent $\nu_\parallel$
is almost the same as that of the square lattice value ($\approx
1.33$) whereas $\nu_\perp$ is higher than that of the square lattice
($\approx 1.12$). Both the exponents are also different form that of
DP model \cite{ebg,dsp1}. The hyperscaling relations $2\delta - 3\gamma
= (d-1)\nu_\perp +\nu_\parallel$ and $(d-d_f)\nu_\perp = \beta =
\delta-2\gamma$ \cite{dsp1} are satisfied marginally: $2\delta -
3\gamma = 2.66 \pm 0.06$ whereas $(d-1)\nu_\perp +\nu_\parallel = 2.59
\pm 0.04$ and $\delta-2\gamma = 0.34 \pm 0.06$ whereas
$(d-d_f)\nu_\perp = 0.28 \pm 0.03$. It is already known that the
hyperscaling is violated in directed percolation \cite{hp}. In the
case of DSP, the hyperscaling relations were found satisfied on the
square lattice whereas on the triangular lattice they are
``marginally'' satisfied. The ratio of the connectivity lengths goes
as $\xi_\parallel/\xi_\perp \sim |p-p_c|^{-\Delta\nu}$ where
$\Delta\nu = \nu_\parallel - \nu_\perp$. For the square lattice,
$\Delta\nu$ is approximately $0.21$ whereas for the triangular
lattice, it is approximately $0.13$. Thus, the clusters are less
anisotropic on the triangular lattice. This is because of more
flexibility given to the spiraling constraint which makes the cluster
not only compact but also less anisotropic.

The values of the critical exponents and fractal dimension obtained in
the above study for the triangular lattice are summarized and compared
with the square lattice data in Table.\ref{table}. The values in the
parenthesis are the suggested rational fractions for the values of the
critical exponents on the square lattice. These rational fractions
satisfy the scaling relations exactly including the hyperscaling
relations. It can be seen that some of the critical exponents and the
fractal dimensions are considerably different on the square and
triangular lattices for the DSP model. According to the theory of
critical phenomena, the values of the critical exponents are
independent of the underlaying lattice structure in the same spatial
dimension.  As a consequence, the systems defined on different
lattices in the same space dimension then belong to the same
universality class. Since the values of the critical exponents of the
DSP model differ on the square and triangular lattices in $2D$, the
DSP model then exhibits a breakdown of universality. This is the first
percolation model which shows breakdown of universality on two
different lattices in the same spatial dimension. It is already seen
in the above discussion that the flexibility in the spiraling
constraint makes the clusters compact and less anisotropic. A possible
reason for different critical behaviour on the square and triangular
lattices may be due to different scaling behaviour of the finite
clusters below percolation threshold on the two lattices. Below $p_c$,
the finite clusters are called lattice animals\cite{lub}. Lattice
animals without any loop are known as lattice trees. Though the spiral
lattice animals have the same scaling form on the square and
triangular lattices, it has been found that the spiral trees (lattice
animals without loops) follow two different scaling relations on the
square and triangular lattices and belong to two different
universality classes\cite{slt}. In the asymptotic $n\rightarrow\infty$
limit, the number of spiral lattice site trees $(a_n)$ of $n$-sites on
the triangular lattice obey the scaling relation given by $a_n \approx
\lambda^{n^\delta} n^{-\theta}$ \cite{slt} whereas on the square
lattice it is given by $a_n \approx \lambda^{n} n^{-\theta}$
\cite{sla}, where $\delta$ and $\theta$ are two exponents and
$\lambda$ is known as the growth parameter. The origin of different
scaling forms for the spiral lattice site trees on the square and
triangular lattices is due to the fact that they can not have
branching on the triangular lattice except at the origin whereas on
the square lattice branching is possible at any point. The radius of
gyration exponent of spiral trees was also found different on the
square ($\nu\approx 0.653$) and triangular lattices ($\nu\approx
0.618$) \cite{slt}. Another lattice statistical model, the spiral
self-avoiding walks (SAW) also exhibit breakdown of universality on
the square and triangular lattices. The asymptotic (large $n$)
behaviour of the number of walks $S_n$ is given by $S_n \approx
An^{-\gamma} \exp(\lambda \sqrt{n})$, where $A=2^{-2} \times 3^{-5/4}
\pi$, $\gamma=7/4$, and $\lambda = 2\pi/\sqrt{3}$ for the square
lattice \cite{ssaw1} and $A = 2^{1/4} \times 3^{-7/4} \pi$,
$\gamma=5/4$, and $\lambda = \pi/\sqrt{2/3}$ for the triangular
lattice \cite{ssaw2}. Notice that the scaling relation for the spiral
lattice site trees is similar to that of the spiral SAWs. It should be
mentioned here that the values of the critical exponents and the
scaling behaviour of the cluster related quantities in the SP model
(percolation in the presence of rotational constraint only) are the
same on the square and triangular lattices and on breakdown of
universality has been observed\cite{spst}. This may be due to the fact
that the spiral lattice trees are minority in number at the
percolation threshold and unable to change the universality class. In
the DSP model, the presence of the directional constraint on top of
the rotational constraint might increase the number of spiral lattice
trees. Maybe, the higher number of tree like structures in the
clusters generated has a non-trivial effect on the critical properties
of the DSP model at the percolation threshold and leads to breakdown
of universality of the critical exponents.

\begin{figure}
\bigskip
\centerline{\hfill \psfig{file=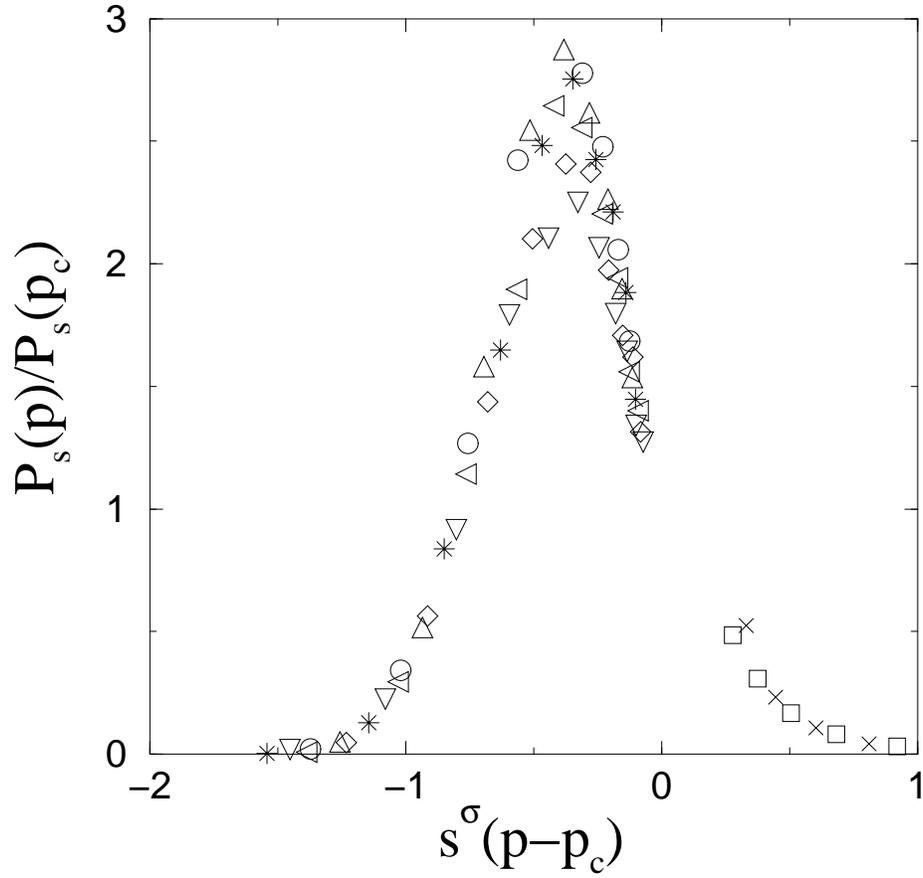,width=0.75\textwidth} \hfill}
\bigskip
\caption{\label{datac} Plot of the scaled cluster size distribution
$P_s(p)/P_s(p_c)$ versus the scaled variable $s^\sigma(p-p_c)$ for
different values of $p$ on the triangular lattice. The value of
$\sigma$ is taken as $\sigma=0.427$. The cluster size $s$ changes from
$64$ to $16384$. The data plotted correspond to $p-p_c= 0.007
(\times)$, $0.005 (\Box)$, $-0.035 (\bigtriangledown)$, $-0.04
(\Diamond)$, $-0.045 (\lhd)$, $-0.05 (\ast)$, $-0.055
(\bigtriangleup)$, $-0.06 (\bigcirc)$. A reasonable data collapse is
observed. }
\end{figure}

Finally, the form of the scaling function $P_s(p) =s^{-\tau+1} {\sf
f}[s^\sigma(p-p_c)]$ is verified. The exponents $\tau$ and $\sigma$
are estimated using the scaling relations $\beta=(\tau-2)/\sigma$,
$\gamma=(3-\tau)/\sigma$, $\delta=(4-\tau)/\sigma$, and
$\eta=(5-\tau)/\sigma$ and following the same technique described in
Ref.\cite{dsp1}. The estimates of $\tau$ and $\sigma$ are obtained as
$\tau= 2.16 \pm 0.02$ and $\sigma= 0.427 \pm 0.003$ respectively. The
errors quoted here are the propagation errors. On the square lattice,
the values of $\tau$ and $\sigma$ were obtained as $\approx 2.16$ and
$0.459$ respectively. Note that, the value of $\tau$ is the same as
that of the square lattice whereas $\sigma$ differs on the two
lattices. The scaling function form is verified through data collapse
by plotting $P_s(p)/P_s(p_c)$ against the scaled variable $s^\sigma
(p-p_c)$ in Fig. \ref{datac}. The cluster size $s$ changes from $64$
to $16384$ and $(p-p_c)$ varies from $0.007$ to $-0.06$. A reasonable
data collapse is observed. The scaling function form is similar to
that of the square lattice. The height of the function remains almost
the same but the width is slightly smaller than that of the square
lattice function. The lesser width of the scaling function $P_s(p)$ is
just a consequence of the lesser value of $\sigma$ on the triangular
lattice.

\section{Conclusion}
The directed spiral site percolation is studied on the triangular
lattice and the results are compared with that of the square
lattice. Clusters on the triangular lattice are found more compact and
less anisotropic than the clusters on the square
lattice. Interestingly, it is also found that the values of the
fractal dimension and some of the critical exponents on the triangular
lattice are significantly different from the square lattice
values. This might be due to different scaling behaviour of some of
the finite clusters below percolation threshold on the two
lattices. As a consequence, the DSP model exhibits a breakdown of
universality between the square and triangular lattices in $2D$. The
values of the critical exponents on both the lattices satisfy the
scaling relations between the moment exponents ($\gamma,\delta,
\eta$). The hyperscaling relations were satisfied on the square
lattice but they are ``marginally'' satisfied on the triangular
lattice. The exponents are not only different on the square and
triangular lattices but also different from other percolation models
like OP, DP and SP. Directed spiral percolation is expected to occur
in disordered systems when both rotational and directional force
fields are present.

\section{acknowledgment}
The authors thank Deepak Dhar for helpful discussions.

\newpage

\vfill

\begin{table}
\begin{tabular}{p{2.4cm}p{2.5cm}p{2cm}p{2cm}p{2cm}p{2cm}p{2cm}}
\hline Lattice Type  & $d_f$ & $\gamma$ & $\delta$ & $\eta$ &
$\nu_\parallel$ & $\nu_\perp$\\
\hline 

Square\cite{dsp1}: & $1.733\pm 0.005$ & $1.85\pm 0.01$ & $4.01\pm
0.04$ & $6.21\pm 0.08$ & $1.33\pm 0.01$ & $1.12\pm 0.03$\\
& $(12/7)$ & $(11/6)$ & $(24/6)$ & $(37/6)$ & $(4/3)$ & $(7/6)$\\
\begin{tabular}{p{2.4cm}p{3cm}}
 & $1.72\pm 0.02$ (FS)
\end{tabular}\\ 
Triangular: & $1.775\pm 0.004$ & $1.98\pm 0.02$ & $4.30\pm 0.04$ &
 $6.66\pm 0.08$ & $1.36\pm 0.02$ & $1.23\pm 0.02$ \\ 
\begin{tabular}{p{2.4cm}p{3cm}}
 & $1.80\pm 0.03$ (FS)
\end{tabular}
 \\ \hline
\end{tabular}
\bigskip
\caption{\label{table} Comparison of the critical exponents and
fractal dimension of the DSP model measured on the square and
triangular lattices. For the square lattice, the values within
parenthesis are the suggested rational fractions for the values of the
critical exponents in Ref.\cite{dsp1} which satisfy the scaling
relations exactly. Some of the critical exponents and the fractal
dimension are significantly different on the two
lattices. }
\end{table}

\vfill

\newpage

\end{document}